\documentclass[conference]{IEEEtran}
\IEEEoverridecommandlockouts
\usepackage{cite}
\usepackage{amsmath,amssymb,amsfonts}
\usepackage{algorithmic}
\usepackage{graphicx}
\usepackage{textcomp}
\usepackage{xcolor}
\usepackage[ruled]{algorithm2e}
\usepackage{adjustbox}
\usepackage{tabularx}
\graphicspath{{./images/}}
\DeclareGraphicsExtensions{.pdf, .eps, .jpg, .png}

\newlength\mylength
\newlength\mylengthb
\usepackage{array}
\newcolumntype{C}[1]{>{\raggedright\arraybackslash}p{#1}}

\def\BibTeX{{\rm B\kern-.05em{\sc i\kern-.025em b}\kern-.08em
    T\kern-.1667em\lower.7ex\hbox{E}\kern-.125emX}}

\begin{document}

\title{Application Inference using Machine Learning based Side Channel Analysis\\
}

\author{\IEEEauthorblockN{ \textsuperscript{1}Nikhil  Chawla, \textsuperscript{1}Arvind Singh, \textsuperscript{2}Monodeep Kar and \textsuperscript{1}Saibal Mukhopadhyay}
\IEEEauthorblockA{\textit{\textsuperscript{1}Dept. of Electrical and Computer Engineering, Georgia Institute of Technology, Atlanta, USA} \\
\textit{\textsuperscript{2}Intel Corporation, Hillsboro, OR, USA} \\
(nchawla6, asingh367)@gatech.edu, monodeep.kar@intel.com, saibal.mukhopadhyay@ece.gatech.edu}
% \and
% \IEEEauthorblockN{2\textsuperscript{nd} Given Name Surname}
% \IEEEauthorblockA{\textit{dept. name of organization (of Aff.)} \\
% \textit{name of organization (of Aff.)}\\
% City, Country \\
% email address}
% \and
% \IEEEauthorblockN{3\textsuperscript{rd} Given Name Surname}
% \IEEEauthorblockA{\textit{dept. name of organization (of Aff.)} \\
% \textit{name of organization (of Aff.)}\\
% City, Country \\
% email address}
% \and
% \IEEEauthorblockN{4\textsuperscript{th} Given Name Surname}
% \IEEEauthorblockA{\textit{dept. name of organization (of Aff.)} \\
% \textit{name of organization (of Aff.)}\\
% City, Country \\
% email address}
}

\maketitle

\begin{abstract}
The proliferation of ubiquitous computing requires energy-efficient as well as secure operation of modern processors. Side channel attacks are becoming a critical threat to security and privacy of devices embedded in modern computing infrastructures. Unintended information leakage via physical signatures such as power consumption, electromagnetic emission (EM) and execution time have emerged as a key security consideration for SoCs. Also, information published on purpose at user privilege level accessible through software interfaces results in software-only attacks. In this paper, we used a supervised learning based approach for inferring applications executing on android platform based on features extracted from EM side-channel emissions and software exposed dynamic voltage frequency scaling (DVFS) states. We highlight the importance of machine learning based approach in utilizing these multi-dimensional features on a complex SoC, against profiling-based approaches. We also show that learning the instantaneous frequency states polled from on-board frequency driver (cpufreq) is adequate to identify a known application and flag potentially malicious unknown application. The experimental results on benchmarking applications running on ARMv8 processor in Snapdragon 820 board demonstrates early detection of these apps, and atleast 85\% accuracy in detecting unknown applications. Overall, the highlight is to utilize a low-complexity path to application inference attacks through learning instantaneous frequency states pattern of CPU core. 
\end{abstract}


\section{Introduction}
Rapid proliferation of technology has resulted in large-scale growth in the smart devices with tremendous computing capabilities to form the interconnected network of things (IoTs). These computing devices generate huge amounts of data, which carry sensitive information like account details, social security number, passwords, PIN etc. This sensitive data is vulnerable to attacks that invade user’s privacy. To start with, applications unintentionally installed on device can be potential malware threats. Similarly, hijacking the control flow at runtime targeted through software vulnerabilities to launch attacks can trigger shell code. Application inferencing is yet another possibility, wherein untrusted application can identify activities on a device through eavesdropping software events hiding in the background. These are few attack scenarios prevalent in smartphones. To defend against these attacks, an application or program profile should be generated. This can be achieved using distinct signatures acquired through numerous side-channel leakage sources. For instance, in hardware, power, EM-emissions, timing and in software through unintended published information from process logs\cite{1}. It can also be attained utilizing performance tools like perf in linux based systems. Generally, profile generation can be based on existing statistical models or it can be closely approximated  using supervised learning algorithms.  

\begin{figure}[!t]
\centering
\includegraphics[width=\columnwidth,trim=0cm 5.8cm 0cm 4cm,clip]{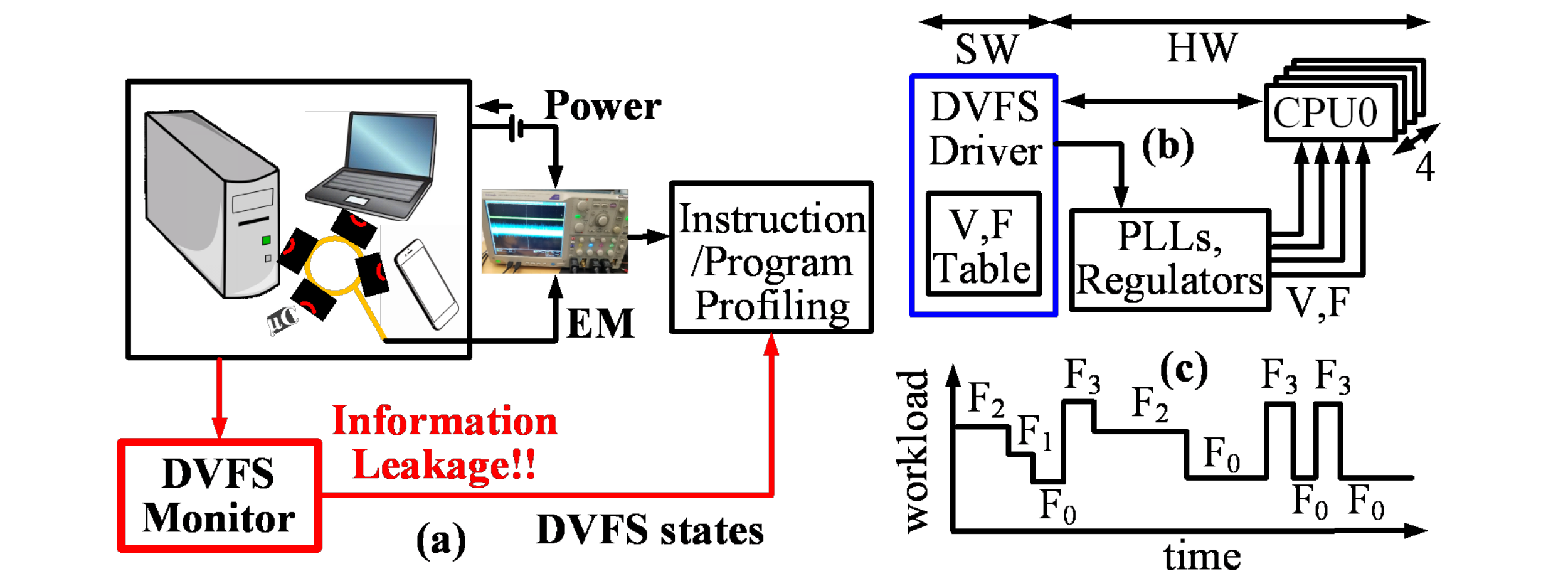}
\label{workload}
\caption{(a) Dynamic voltage frequency scaling algorithms can be a critical side channel for analyzing program behavior, (b) hardware-software co-optimization approach for power management in modern SoCs, and (c) dynamic management of frequency (and voltage) under varying workloads for a quad-core system.}
\label{workload_fig}
\end{figure}

Profiling based side channel attacks comprise of Template Attack (TA) or Stochastic Approach (SA). In profiling-based side-channel attacks adversary models the leakage behavior of device, and later infers secret based on these models. Template attacks (TA) are a special class of profiling-based attacks which are shown to recover cryptographic keys with fewer traces in both hardware and software implementations\cite{2}. Similarly, templates based on software events have been utilized to characterize applications executing on devices. But, template attacks are effective approaches for well-understood devices and its complexity increases for modern System on Chips (SoC’s)\cite{3}. For instance, software events extrapolated from high performance computing-performance monitors (HPC-PM) vary from one kind of architecture to another due to the variation in hardware organizations thereby requiring knowledge of the architecture.  To resolve this problem, machine learning approaches are a suggestive alternative to side channel analysis. Given complex dataset, ML algorithms perform classification or prediction based on experience learned from the examples. In contrast to TA, machine learning based attacks are promising in black box settings, with only limited understanding of the target implementation\cite{3}. Authors in \cite{4} have shown ML-based approach can be more successful than TA, if proper algorithm and tuning parameters are selected. Moreover, the problem of high dimensional side-channel traces can be effectively managed with ML-based approaches than TA, which are best model if fewer Points of Interest (POI’s) could be identified with most information\cite{3}. Several works utilizing ML models have shown to outperform profiling-based attacks like TA. An attack targeting individual bits of TripleDES using Random Forest (RF), Support Vector Machines (SVM) and Self-Organizing Maps (SOM) has been shown to outperform TA\cite{5}. Similarly, Lerman et al. showed an attack on AES with Rotating S-boxes Masking (RSM) countermeasure using SVM and RF and it outperformed TA and SA\cite{6}.

 This paper focuses on performing application inferencing on Snapdragon 820 Quad core processor based on ML models which are trained on features extracted from side-channel emissions of labelled dataset of applications(legitimate). Effect of decision threshold of the trained model on accuracy of detecting an application unknown to the model is also analyzed. The classification accuracies are evaluated through supervised learning models - K-Nearest Neighbors (KNN), SVM and RF. Also, this paper, experimentally studies the impact of EM-emissions and for the first time DVFS as a source of side-channel information leakage (Fig. 1). The paper makes following key contributions:

\begin{itemize}
\item We develop supervised machine learning (ML) based classification models to exploit the relationship between time-varying EM-emissions and DVFS states with applications’ characteristics to identify applications running on processor. 
\item We introduced transitions of DVFS states in time-domain as a side channel information leakage path. We show that change in DVFS states during an application’s runtime can be used to identify applications.  
\item We show that the developed models can be used to classify known programs and detect unknown programs at lower computational and memory cost, but requiring more time compared to EM-based profiling. 
\end{itemize}

We demonstrate the side-channel information leakage via EM-emissions and DVFS through hardware measurements on Snapdragon 820 development kit with ARMv8 processor running Android benchmark applications. The experimental results show close to 94\%(EM) and 80\%(DVFS) accuracy in classifying ‘known’ applications and over 85\% accuracy in detecting 'unknown' applications.  

\section{Background}

\subsection{Electromagnetic (EM) Side Channel}

 Side channel emissions originating as result of electromagnetic radiations from device carry sensitive information about data being processed. The rise of emissions is reflection of sudden surge in current through power rails as CMOS devices switch states and through interconnects and package. In the past, emissions through EM side-channel are utilized for revealing cryptographic keys based on correlation or differential analysis models. Many other complex attacks models have also shown to compromise security of wide domain of computing devices which includes smartphones, smart-cards, FPGA and personal computers. In another setting, electromagnetic emissions(EM) have shown to reveal information about kind of program running on device or can be used as a defense mechanism to identify malicious code, and even perform reverse engineering attacks. The existing approaches to EM side-channel analysis requires physical access or close-proximity to the device to measure time-dependent EM signature.

\subsection{DVFS-based Power Management}

The energy dissipation is a major constraint to IoT devices; hence, processors for IoT/mobile platforms incorporates various dynamic power management mechanisms. In particular, dynamic voltage and frequency scaling (DVFS), has become an integral part of modern processors to improve energy efficiency, increase battery life, and manage thermal effects. With DVFS, the supply voltage and operating frequency are scaled with respect to varying workloads of the target system. Significant efforts have been devoted in past decades in developing DVFS algorithms. Moreover, to enhance energy-efficiency of processors, there has been a constant push to enable software-based DVFS control. Consequently, modern processors have multiple embedded DVFS algorithms available as drivers (referred to as the governors) that users can exercise and monitor the voltage/frequency states of the processors. Different DVFS algorithms are designed based on certain policies like workload sampling interval, frequency upscaling, and downscaling factors etc. The CPU utilization-based decision to down-scale or up-scale frequency, is dependent on busy time of CPU, which in turn is application dependent.  A skilled user can also implement their own DVFS algorithms to augment/replace existing governors and control the voltage/frequency states of the processors. Thus, in nutshell frequency states of all cores are continuously updated by cpufreq driver in Linux kernel and no special privileges are required for user access. This DVFS frequency states creates a software-based path to side-channel information leakage.

\begin{figure*}[!t]
\centering
\includegraphics[width=\textwidth,trim={2cm 3.9cm 2cm  3.9cm},clip]{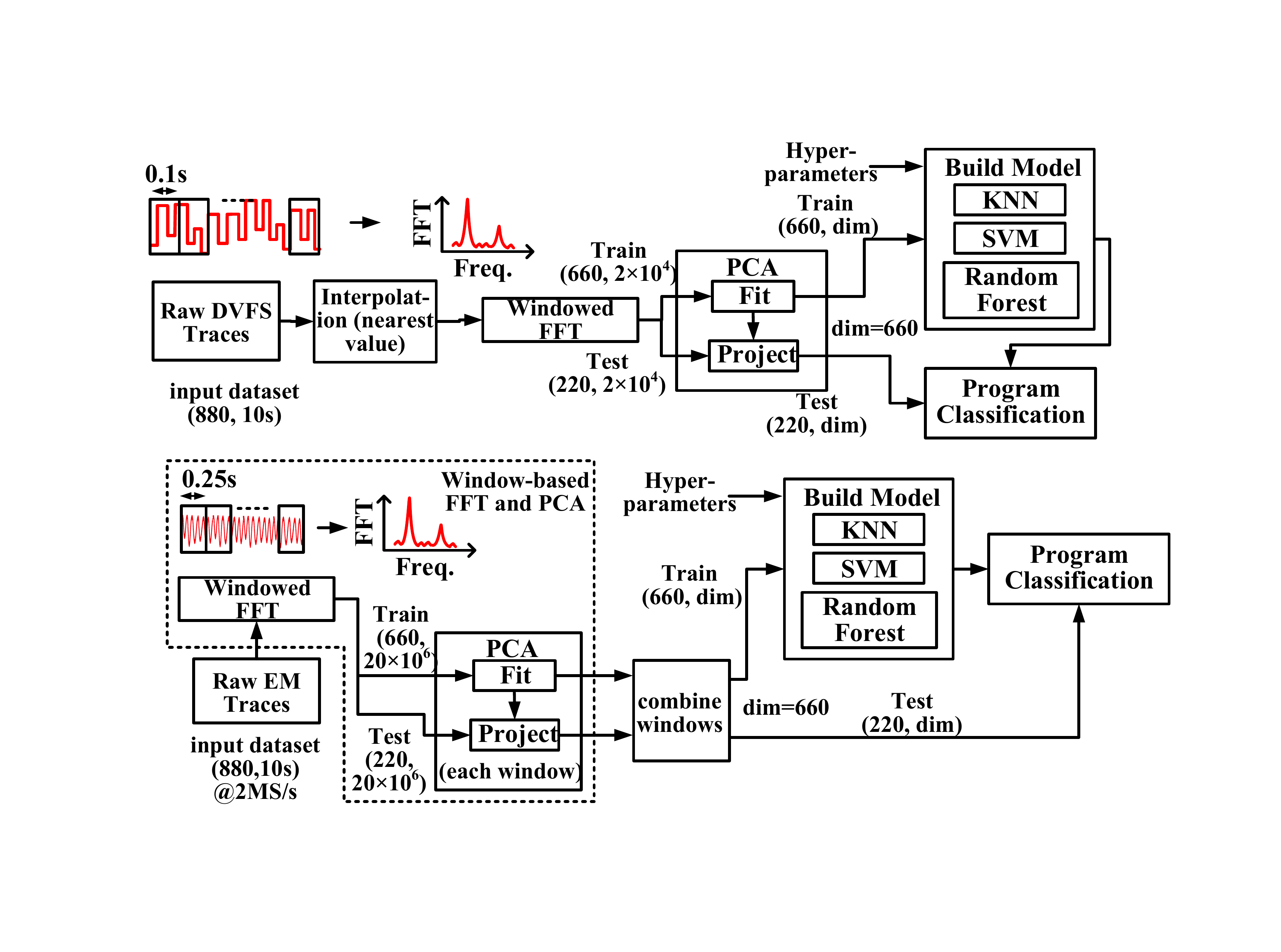}
\label{methodology}
\caption{Pre-processing and feature extraction from the raw measured traces for (a) DVFS and (b) EM signatures followed by building ML-based models and validation.}
%\label{workload}
\end{figure*}

\subsection{Related Work}
Many of the past efforts have focused on EM based side-channel analysis. Nazari et. al. and Callan et. al. presented EM-based acquisition and analysis flow to detect code change injected by an adversary \cite{7,8}. ML-models were used to learn features extracted from power side channel emissions to create a model for detecting malware in medical devices\cite{9}. Similarly, a security monitor for control flow integrity of programs executing on industrial PLC has been demonstrated using LSTM network based on features derived from EM-emissions \cite{10}.

DVFS based power management has been explored extensively across all platforms but only recently researchers have started exploring the interactions of DVFS and security \cite{11,12,13,14}. Yang et. al. has demonstrated use of DVFS as a countermeasure to power side channel attack on encryption engines \cite{11}. More recently. A. Singh et. al. has demonstrated use of fast DVFS enabled by on-chip regulator and adaptive clocking to deter extraction of encryption key in hardware accelerators\cite{12,13}. More recently, Tang et. al. presented a CLKSCREW methodology which exploits the flaws in power management techniques of an ARMv7 processor \cite{14}. By performing unconstrained overclocking/under-volting, authors could inject faults during encryption and successfully recover the secret key.  

Both profiling based and ML based techniques have been
utilized for application inferencing. For instance, to protect the devices against malwares, authors have demonstrated malware detection based on HPCs by selectively choosing application specific hardware and software events \cite{15}. Similar approach, is shown by R. Spritzer et.al. using selective information from
process logs that form a strong correlation for same application in order to form templates and utilizing dynamic time warping(DTW) for application inference. They have shown upto 96\% accuracy in identifying 100 android applications \cite{16}. ML-based approaches have been used to identify malware based on features extracted from power-emissions\cite{11} and HPCs\cite{17}.
Software exposed information can be used to mount inference attacks. Using /proc/$<$pid$>$/statm along with number of context switches, is shown to infer a visited-webpages by a user\cite{18}. Similarly, size of the memory footprint of specific applications in /proc/$<$pid$>$/statm is used to infer the user interface. More recently, activity transitions of an application are inferred using runtime memory statistics.\cite{19} Moreover, identifying which application is running can help in launching specific attacks. For instance, an app running in the background to identify application which require login credentials, can execute phishing-based attacks\cite{19} to steal login credentials.

\section{Methodology}
Our experimental setup includes a Quad-core Snapdragon 820 SoC which has 2 slowly clocked and 2 fast clocked Kyro cores and hosts Android 7.0 Nougat OS. These cores form two clusters and with respective DVFS states. Since, we are profiling applications based on DVFS states running in multi-core platform and power domains of these multi-cores are different, an application’s workload behavior might influence the DVFS states of both the clusters. Therefore, DVFS states is collected from both these clusters.

\begin{figure*}[!t]
\centering
\includegraphics[width=\textwidth,trim={0cm 0cm 0 0cm},clip]{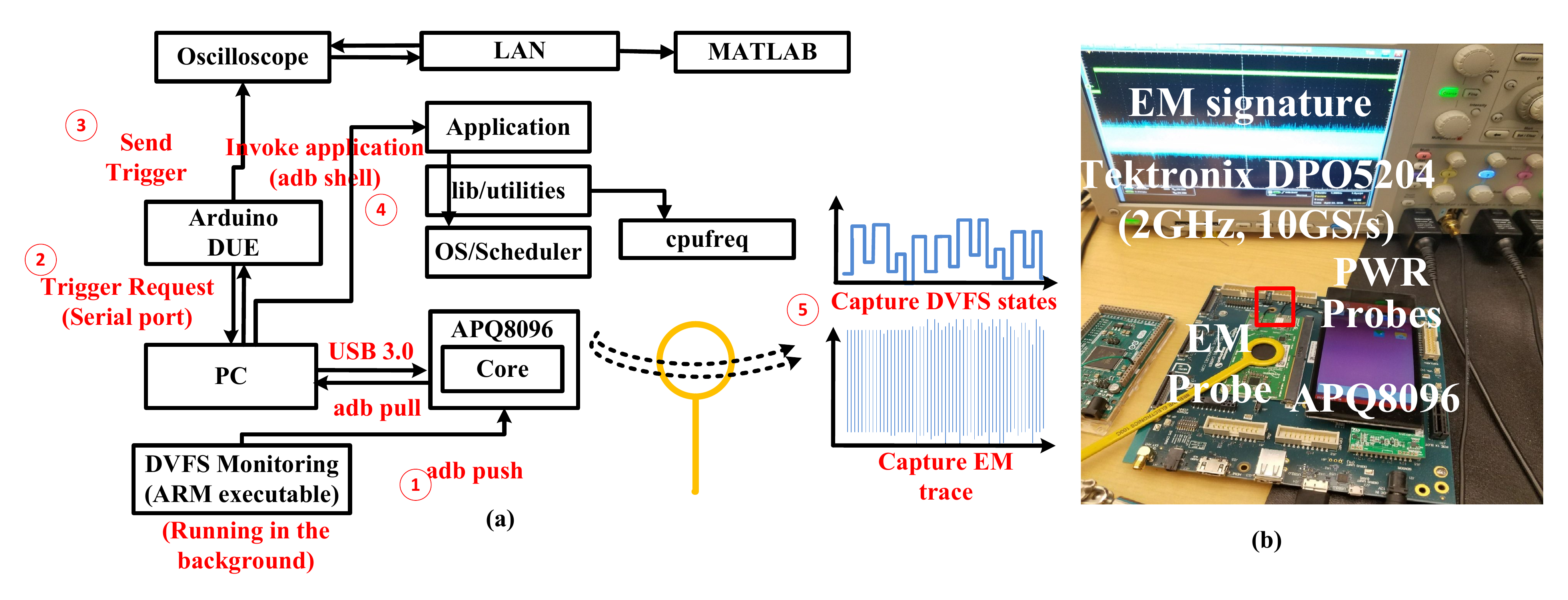}
\label{measurement_setup}
\caption{(a) Measurement Setup details to capture DVFS and EM-signatures (b) Open-Q APQ8096 System-on-Module development platform for characterization.}
\end{figure*}

\subsection{Model Building and Validation with Machine Learning (ML) Algorithms}
We have explored supervised machine learning models for extracting features from time-series of DVFS states and EM-signatures for applications inferencing. ML models that work effectively for high-dimensional signatures derived from EM-emissions have been realized in the past. Here, we analyze ML model selectivity based on DVFS dataset. The learned ML models are used to classify an application to one of the known ones or as an unknown one. The details of our approach are discussed below:

\subsubsection{Training}: To build ML models for the program runtime behavior using EM-emissions and DVFS states, multiple signatures were captured for the same program to ensure enough observations for each program to the effect of noise generated by background activity. Number of records varies with each observation as result of different polling delays.  Therefore, the measured traces are interpolated to have a uniform time interval. The polling frequency encounters 0.5 ms delay, and two clusters are simultaneously polled. The measured traces are partitioned in train data and test data (75\%/25\%).

\subsubsection{Feature Extraction}: For DVFS states, we used time-series values as the features. For EM-side channel we used spectral components. We capture two time-series corresponding to two clusters for DVFS data. Instead of choosing the exact frequency values, we chose frequency indexes (0 to 39) to avoid need for normalization of data as some machine learning algorithms (e.g., support vector machines) do not perform well if the input data has large values. The frequency range of core varies from 307200 to 1593600 KHz with 14 states in between for 2 cores, and 307200 to 2150400 KHz with 16 states in between for other 2 cores. We also explored features extracted from frequency domain, which in fact showed improved classification results. Frequency states from both the cores are appended to form the feature vector. Additionally, since number of time samples in the input dataset are very large, we performed principal component analysis (PCA) to bring down the dimensions within 660. PCA reduces the dimensionality keeping the variance to as much as possible to maximum. We have evaluated relation of Principal Components required against variance in the dataset. 

\subsubsection{Machine Learning Algorithms}: To learn program runtime characteristics from DVFS states, we chose supervised machine learning algorithms suitable for varying type of datasets. Both our training and testing datasets vary a lot in terms of dimensionality, number of observations and noise characteristics which in turn depend on the selected DVFS governor, application that is being run, the sampling speed at which the DVFS monitor is capturing the frequency states and OS essential apps/services running in the background. Here we describe the chosen algorithms and the reasons for their selection and our expectation with respect to this specific classification problem.  

\underline{\textit{K-Nearest Neighbors (KNN)}}: KNN algorithm can classify datasets with linear or non-linear distributions. It just assigns labels to training examples without building any models. For inference, it looks at the ‘k’ nearest neighbors based on the distance defined as a function of features. Both distance metric and number of neighbors are input to the algorithm and can be tuned for a given dataset. KNN performs well with large number of observations in low-dimensional feature space. Since, the effective dimensionality of our DVFS dataset depends on the real-time workload conditions and the selected DVFS governor, KNN is a good option for applications having mostly constant workloads. However, KNN tends to overfit with respect to noisy data or some bad features so its performance may not be good in presence of lot of background activity. In our experiment, we vary ‘k’ from 1 to 20 to find the optimal value. Our distance metric is Euclidean. 

\underline{\textit{Support Vector Machines (SVM)}}: Like KNN, SVM can also classify linearly or non-linearly distributed datasets with a proper kernel (linear, poly, radial basis function – RBF). SVM finds and hyperplane to separate out the given classes. SVM can be tuned with respect to its parameters (C, Gamma) and the type of kernel. For noisy as well as high-dimensional data, SVM tends to outperform other ML algorithms. One major drawback with SVM is its training runtime. To reduce the training runtime, dimensionality reduction with principal component analysis (PCA) is performed before model building step. The DVFS frequency states have linear dependence on the program workload conditions, therefore, a linear kernel is expected to perform better for SVM based learning and validation.

\underline{\textit{Random Forests}}: Random forest, being an ensemble learning algorithm (collection of several weak classifiers), performs classification by using multiple randomized decision trees (DT). Like SVM, they avoid overfitting with DT randomization. However, unlike SVM which was mainly designed for binary classification problems, Random forest applies to multi-class problem, very useful for our 10-class datasets. Additionally, Random forest works with numerical and categorical features suitable for our learning problems where features are the frequency states. Due to randomization, we take the average of 10 runs for random forest-based learning. In our analysis we varied n\textunderscore{estimators} (an input to Random Forest algorithm), from 10 to 100 and found the optimal to be 40.

\begin{table}[htbp]
\caption{Hyper-parameters tuned/explored for machine learning algorithms employed in this work}
\begin{center}
\begin{tabular}{C{\mylengthb}C{\mylengthb}}
%\begin{tabular}{|c|c|}
\hline
\textbf{ML Algorithm} & \textbf{Hyper\textunderscore{parameters}} \\
\hline
KNN & num\textunderscore{neighbors}=1 to 20 \\
%\hline
SVM & C=1, Gamma=auto, linear kernel \\
%\hline
RF & num\textunderscore{estimators}=40 \\
\hline
\end{tabular}
\label{tab1}
\end{center}
\end{table}

Table 1 lists the hyper-parameters chosen/tuned for our experiments. We will discuss our findings with different ML models in the results and analysis section. Both of our datasets are noisy: DVFS traces are noisy because of background activity and EM traces because of measurement, environment. KNN tend to overfit in presence of noise while SVM performs well.

\subsection{Overall Approach}
 Fig. 2 describes the overall flow to monitor the DVFS states and EM-emissions followed by subsequent analysis to build machine learning models and classify known programs and to detect unknown programs. The frequency states act as unintentional source of information leakage from software and are recorded using a script running in the background, that reads current operating frequency of cluster from a cpufreq file system.To avoid noise being generated from the other cores, no other application is running in the background, apart from essential system applications/services. Applications running in parallel with profiled application will generate a cumulative utilization pattern that distinctly don’t represent leakage from DVFS states of the profiled application. Moreover, performance critical application executing in parallel can even mask the leakage of profiled application because of substantial contribution to the overall utilization of the core. Therefore, this methodology only profiles single application. In Acquisition phase, benchmarks were executed, 40 traces for each application were gathered for a runtime of 10 seconds. In the next section, we will describe the experimental setup, post-processing steps and explored machine learning algorithms in more details.
 
\section{Experimental Verification}
\subsection{Measurement Setup}
An integrated mechanism to simultaneously capture DVFS states and EM side channel signatures is developed, which is followed by post-processing steps performed offline. The Benchmarking APKs are installed on Snapdragon 820 development kit assembled by Intrinsyc Open-Q 820 (APQ8096) System on Module (SOM)[Figure 3(a,b)] which hosts an Android (v7.0) OS. A separate LCD screen is installed for user interface. The key components in the acquisition process are detailed below

\subsubsection{DVFS Trace Acquisition}: The interface between the development kit and desktop is established through Android Debug(ADB) Interface. Two separate compiled binaries for polling the frequency states of individual core from cpufreq files is loaded onto Android device using \textit{adb push} command. An algorithm highlighting the steps included in DVFS state monitoring script is shown in Algorithm 1.  To successfully capture all frequency state transitions, polling frequency should exceed frequency update rate in cpufreq module. Built-in governor samples the workload in the order of (20-80) ms. Total polling delay per acquisition is 0.5 ms. The applications are profiled for duration of 10s.

\subsubsection{EM Trace Acquisition}
The DVFS states are monitored and stored through a script. Hence, DVFS does not require any additional acquisition instrument. However, EM-trace capture has additional constraints for acquisition. An antenna-based probe, which is used to capture these EM signatures has to be carefully chosen. We have used a large loop antenna probe (Beehive Corp., 1” tip diameter and 0.85” loop diameter, 50MHz 3dB bandwidth, 50Ω termination). Further, for EM-trace, we set the sampling frequency at 2MHz to capture 10s long experiment. A finite buffer reading time is required for oscilloscope capture which is included in shell script for reliable EM-trace acquisition. 

\begin{algorithm}[t]
\DontPrintSemicolon

pid $\xleftarrow[]{}$ processID(pkgname(app))

state $\xleftarrow[]{}$ read(‘/proc/’pid’/status’)\\  
\While {State(pid) is Running}{
sTime $\xleftarrow[]{}$ Time (currentTime) \\
frequency $\xleftarrow[]{}$ read(‘/sys/devices/system/    \\    cpu/cpuX/cpufreq/scaling\textunderscore{cur}\textunderscore{freq}) \\
eTime $\xleftarrow[]{}$ Time (currentTime) \\
writeFile (sTime, eTime, frequency)
}

\caption{DVFS Monitoring Algorithm}
\label{algo1}
\end{algorithm}

\begin{figure*}[!t]
\centering
\includegraphics[width=\textwidth,trim={4.5cm 8.0cm 5cm 6.5cm},clip]{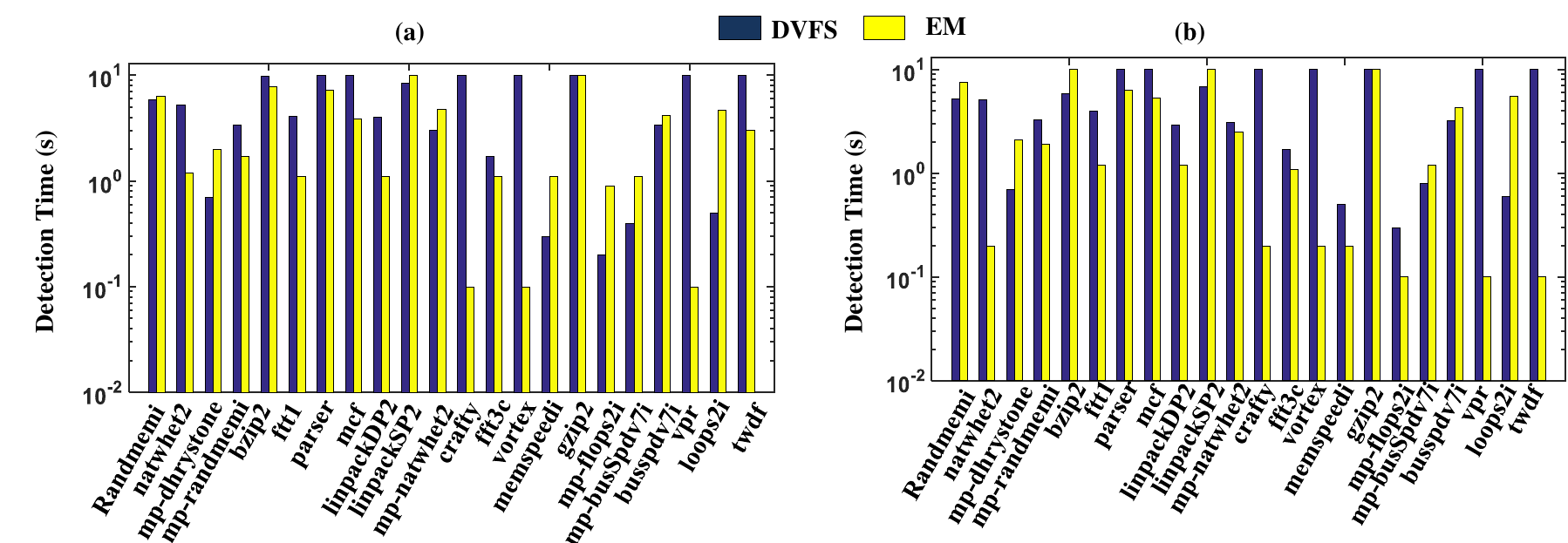}
\label{images/fig1}
\caption{Detection time for applications with (a) SVM inference model and (b) RF inference model using features from DVFS and EM side channel}
\end{figure*}

\subsubsection{Trigger, Log and Fetch}: Trigger is required to synchronize EM and DVFS measurements. 
A trigger signal is generated from Arduino board which handshakes with script running on PC. This trigger signal is used by oscilloscope to capture and store the EM signature for each program. Arduino sends an acknowledgement signal to PC  when it successfully receives command to generate trigger signal. On receipt of Acknowledgement from Arduino, script on PC starts the benchmarking application.

Once the application is triggered, the DVFS monitoring script logs the frequency states and time stamp for observation period of 10s. Finally, the log is saved onto the internal memory of the chip, and process is repeated until all applications with multiple measurements are profiled. During the Fetch phase, saved files are transferred from development kit to an offline environment using \textit{adb pull} command. This is followed by pre-processing steps and Machine Learning training and inference steps.

\begin{table}[htbp]
\caption{Classification Accuracy for DVFS and EM side Channel}
\begin{center}
% \begin{tabularx}{\columnwidth}{X|X|X|X}
\begin{tabular}{C{\mylength}C{\mylength}C{\mylength}C{\mylength}}
\hline
\textbf{DVFS/EM} & \textbf{KNN} & \textbf{SVM} & \textbf{RF} \\
\hline
DVFS (Time) & 62.72 & 67.72 & 76.68 \\
% \hline
DVFS (Freq) & 64.54 & 75.45 & 80.09 \\
% \hline
EM (Freq) & 86.81 & 94.54 & 91.99 \\
\hline
\end{tabular}
% \end{tabularx}
\label{tab2}
\end{center}
\end{table}
\subsection{Application Classification}

\subsubsection{EM side-channel}
Features extracted from EM side-channel emissions are evaluated in frequency domain using window-based FFT with 200 ms window (100 windows in total) [Fig. 2(b)]. Since EM signatures are sampled at very fast rate (2MS/s), we have 20 million-time samples (frequency components in per EM signature). To reduce the dimensionality, window-based PCA is performed and 660 components are retained in total (1-17 in each window). However, EM-SCA is computationally more expensive as well as requires large memory to perform PCA. Table 2 shows the EM-SCA classification accuracies with all ML models. A comparatively lower accuracy is observed with KNN. It is not optimal due to noise in EM measurements coupled from multiple sources. On the other hand, SVM and RF performs better but we do not observe 100\% accuracy with EM-emissions.

\subsubsection{DVFS}
Machine learning models are trained and validated with DVFS signatures measured for all applications (40 signatures per application, 880 signatures in total). The measured data is divided into 660/220 signatures (75\%/25\% split) for training and validation phases. Fig. 2(a) describes the pre-processing steps needed to align the DVFS traces and reduce dimensionality from $2*10^4$ to within 660 principal components using principal component analysis. Feature Extraction was performed on each window separately and finally all features are combined to form a reduced dimension feature vector. We have utilized both time-domain and frequency-domain features for classifying applications. Table 2 shows cumulative accuracies for test data with respect to 3 different machine learning models (KNN, SVM and Random Forest). In general, classification results based on frequency domain features have higher accuracies in comparison to time-domain features with all evaluated machine learning models. On comparing classification accuracies with different ML models, KNN does not perform optimal fitting. It can be concluded it is no best algorithm when data is noisy, or features are not consistent. But, SVM and RF performs better with the dataset, showing 75.45\% and 80.09\% accuracy respectively.

\begin{figure}[!t]
\centering
\includegraphics[width=\columnwidth,trim={6cm 7cm 5cm 5cm},clip]{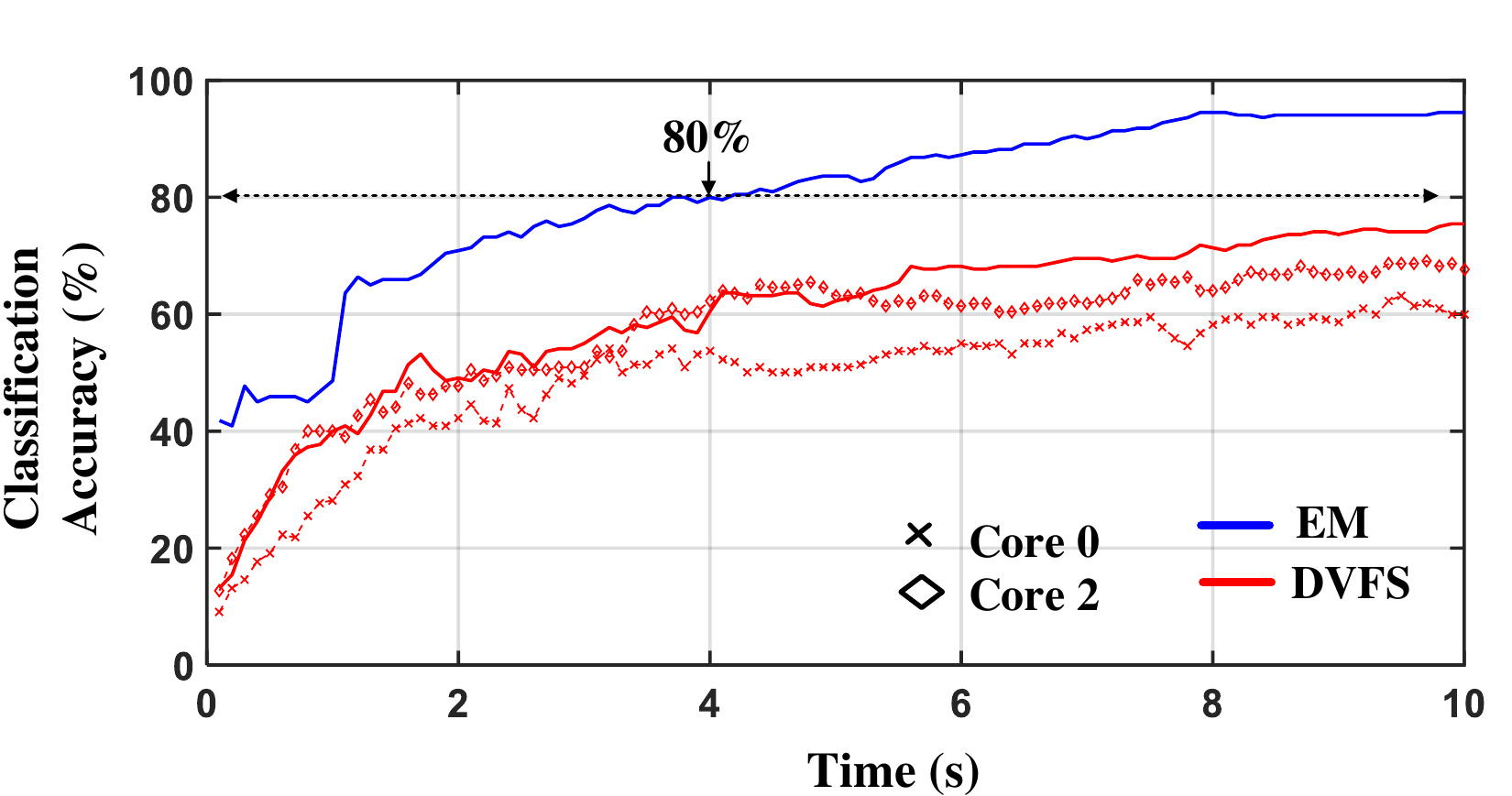}
\label{images/fig2}
\caption{Cumulative Detection time for all applications with SVM inference model using EM-features and DVFS features from high performance, energy efficient core.}
\end{figure}

\subsection{Application Detection}
Classification Accuracies is not the best metric for comparing information leakage from DVFS and EM-SCA. Therefore, we study the latency associated with detecting (classifying) an application with DVFS signatures and compare that with EM-SCA based detection. Detection latency is estimated using window-based classification of DVFS or EM signatures. Window size is chosen based on CPU load evaluated by the governor (order of milliseconds). Data is incremented in steps of window size. As expected, the classification accuracy improves as data from more number of windows are used for classification. The time required for 80\% classification for each application is derived which is multiple of number of windows. The steps involved in calculating detection time is listed in Algorithm 2. The detection time for different benchmark application based on Frequency domain EM-SCA and DVFS is shown in Fig. 4(a,b). Cumulative classification accuracies with KNN is lower in comparison to SVM and RF as shown in Table 2,therefore detection time is evaluated using these two ML algorithms.

\subsubsection{EM side-channel}
Fig. 4 depicts detection latency for applications. Many applications show detection latency of 10s which indicates failure in classifying particular application with 100\% confidence. As EM emissions integrate leakages from multiple sources on-chip, which contributes to noise and reduces EM signal, thereby selected features cannot be segregated. The time-dependence of classification accuracy is  depicted in Fig. 5. It shows gradual increase in cumulative accuracy with 80\% accuracy achieved in 4s time.

\subsubsection{DVFS}
Application detection based on frequency domain DVFS signatures have experimentally shown lower detection latency(average). Though some applications are classified faster compared to EM-SCA as shown in Fig. 4(a, b), overall cumulative accuracy variation over time is slower as shown in Fig5. We have analyzed the impact of selecting features from multicore in comparison to single core through classification accuracy or detection latency as shown in Fig. 5. The detection time drops in case if features are  extracted from either energy-efficient or high-performance. Core migrations effects can result in applications switching between cores based on workload requirements of applications, therefore multi-dimensional feature combined from both cores shows improved classification as well as detection accuracies.

\begin{algorithm}[t]
\DontPrintSemicolon
nApps : Number of applications \\
train\textunderscore{size} : Training data size \\
test\textunderscore{size} : Test data size \\
data : FFT (DVFS time series) \\
nWin : Number of windows  \\
n\textunderscore{components} : Number of principal components \\ 
max\textunderscore{detection}\textunderscore{time} : Maximum detection time \\
\ForAll{apps}{
\ForAll {nWin}{
train\textunderscore{data} $\xleftarrow[]{}$data(nApps,train\textunderscore{size},\\ min\textunderscore{window}\textunderscore{size},windowNumber) \\
test\textunderscore{data} $\xleftarrow[]{}$data(nApps,test\textunderscore{size},\\ min\textunderscore{window}\textunderscore{size},windowNumber) \\
data $\xleftarrow[]{}$ windowedPCA(train\textunderscore{data}, test\textunderscore{data},\\ train\textunderscore{label}, test\textunderscore{label}, n\textunderscore{components})\\
prediction\textunderscore{scores} $\xleftarrow[]{}$ SVM (train\textunderscore{data}, test\textunderscore{data},\\ hyperparamters, train\textunderscore{labels}, test\textunderscore{labels}) \\
accuracy[apps] $\xleftarrow[]{}$ Time (prediction\textunderscore{scores})\\
}}
\ForAll {apps}{
 \uIf{accuracy[apps] \textbf{greater or equals} 80}
 {
    index $\xleftarrow[]{}$ FindIndex(accuracy[apps]  \textbf{equals} 100) \\
    detection time[apps] $\xleftarrow[]{}$
    index * min\textunderscore{window}\textunderscore{time} \\
 }
  \Else{
    detection time[apps] $\xleftarrow[]{}$ max\textunderscore{detection}\textunderscore{time} \\
  }
}
\caption{Application Detection}
\label{algo}
\end{algorithm}

\subsection{Detecting Unknown Applications}
The ML models were trained to learn features of legitimate applications installed on mobile device. This trained model can be used to flag a potentially malicious or application not known to user. The trained model with known applications dataset is applied to a test dataset of some of the unknown applications.We have utilized MiBench applications for the testing, with each program generating 40 signatures during testing. To detect unknown class, we define a decision threshold to classify the unseen data in the unknown program class only when probability for all existing classes less than decision threshold. The degradation in classification performance is measured using precision [True Positives / (True Positives + False Positives)] and recall [True Positives / (True Positives + False Negatives)].

Fig. 6 shows that classification accuracy of unknown apps is atleast 85\% and 88\% for DVFS and EM side channel respectively evaluated with SVM and RF inference models. SVM model shows higher accuracy for unknown application and performs comparatively better with  EM side channel features. RF inference model on DVFS features of test set classify unknown applications with 94\% with known test accuracy reduced to 63\% (more false negatives) at higher threshold values. At lower threshold values, accuracy of unknown application drops to 85\% and 87\% but at higher accuracy of known class (lower false negatives).   

\begin{figure}[!t]
\centering
\includegraphics[width=\columnwidth,trim={5cm 6.5cm 5cm 5cm},clip]{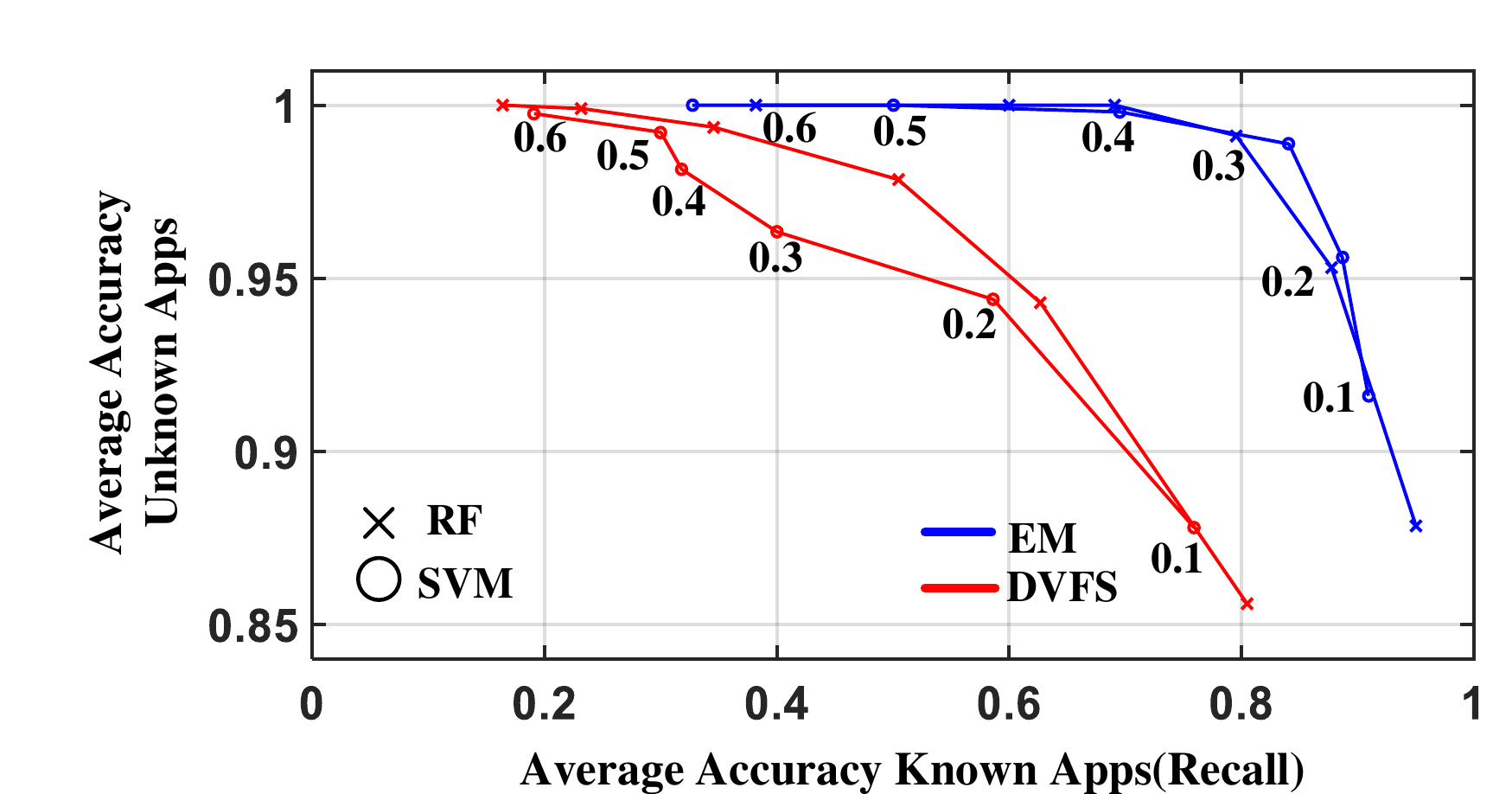}
\label{images/fig3}
\caption{Average Accuracy for Known and Unknown applications, at different decision thresholds with SVM and RF trained model using EM-emissions and DVFS spectral features.}
%\label{workload}
\end{figure}

\section{Conclusion}
This work experimentally demonstrated the feasibility of identifying applications running on a device and detecting unknown application with high accuracy utilizing a machine learning based approach based on EM-side channel and software exposed DVFS information on Snapdragon 820 processor. Although, impact of DVFS states on evading information leakage or promoting other side-channels have been studied in the past, we have shown that information harnessed from these states is useful for identifying applications with less complexity. The experimental results show unknown applications can be identified with atleast 85\% accuracy for maximum achievable accuracy of 94\% for known applications on the analyzed dataset. Application specific DVFS features can classify applications with 80\% accuracy but cannot provide better detection than EM-side channel features. On the other hand, application inference using EM side-channel is more computationally intensive and prone to measurement noise.

\bibliographystyle{./IEEEtran}
\bibliography{ref}
\vspace{12pt}

\end{document}